\documentclass[letterpaper, 10 pt, conference]{ieeeconf}  

\IEEEoverridecommandlockouts                              

\usepackage{graphics} 
\usepackage{epsfig} 
\usepackage{mathptmx} 
\usepackage{times} 
\usepackage{amsmath} 
\usepackage{amssymb}  
\usepackage{cite}
\usepackage{algorithmic}
\usepackage[utf8]{inputenc}
\usepackage[english]{babel}
\usepackage{xcolor}
\usepackage{array}
\usepackage{booktabs,multirow,threeparttable}
\usepackage{comment}

\title{\LARGE \bf
VibroWeight: Simulating Weight and Center of Gravity Changes of Objects in Virtual Reality for Enhanced Realism
}

\author{Xian~Wang$^{1}$, Diego~Monteiro$^{2}$, Lik-Hang~Lee$^{3}$, Pan~Hui$^{1}$, and~Hai-Ning~Liang$^{4,^*}$
\thanks{$^{1}$X. Wang and P. Hui are with the Hong Kong University of  Science and Technology, Hong Kong, China.}
\thanks{$^{2}$D. Monteiro is with Birmingham City University, Birmingham, UK.}
\thanks{$^{3}$L.-H. Lee is with Korea Advanced Institute of Science and Technology (KAIST), Daejeon, South Korea.}
\thanks{$^{4}$H.-N. Liang is with Xi'an Jiaotong-Liverpool University, Suzhou, China.}
\thanks{$^{*}$Corresponding author ({\tt\small haining.liang@xjtlu.edu.cn})}
}

\begin{document}

\maketitle
\thispagestyle{empty}
\pagestyle{empty}

\begin{abstract}

Haptic feedback in virtual reality (VR) allows users to perceive the physical properties of virtual objects (e.g., their weight and motion patterns). However, the lack of haptic sensations deteriorates users' immersion and overall experience. In this work, we designed and implemented a low-cost hardware prototype with liquid metal, VibroWeight, which can work in complementarity with commercial VR handheld controllers. VibroWeight is characterized by bimodal feedback cues in VR, driven by adaptive absolute mass (weights) and gravity shift. To our knowledge, liquid metal is used in a VR haptic device for the first time. Our 29 participants show that VibroWeight delivers significantly better VR experiences in realism and comfort.

\end{abstract}

\section{INTRODUCTION}

Virtual reality (VR) head-mounted displays (HMDs) have gained increasing popularity~\cite{IEEEVR2020-ConsumerHeadsets}. Although users can receive enriched immersive visual experiences in VR, now even at their homes or offices, current VR devices only support simple vibration haptic feedback, limiting users' sense of presence and their immersion in virtual environments (VEs) ~\cite{carlos-eics2021}. Weight properties of virtual objects are crucial to achieving a greater sense of presence, as they allow these objects to be felt just like their physical counterparts~\cite{monteiro2021ICMI}. Not sensing weight properties could negatively impact the sense of presence, primarily due to the absence of muscle stress~\cite{muscle-haptic}. Researchers have proposed several solutions to leverage vibrations and pseudo-haptic to emulate the weight of virtual objects~\cite{hirao2020comparing,van2019visual,samad2019pseudo}. However, neither simple vibration feedback nor pseudo-haptic approaches can sufficiently simulate the weight properties of virtual objects with the necessary level of accuracy and realism. As such, when experienced by users, the interaction process run into the risk of producing sensory conflicts which can degrade user performance and experience~\cite{zenner2017shifty,carlos-eics2021}. Therefore, it is worthwhile to investigate alternative interactive haptic devices that can simulate the weight of virtual objects in VR, including shifts in their center of gravity.  

In this research, we have designed and implemented a hardware prototype, VibroWeight, that enables users to sense the weight properties of objects in VR. VibroWeight offers detailed granularity for VR objects with various weight properties and dynamic changes of manipulated objects, such as (1) varied weightings of rocks with different sizes; (2) a water spray gun with frequent changes in its weight; and (3) the shift in the center of gravity of a wobbling stick or other cylinder-shaped objects. VibroWeight is designed to be easily integrated into commercially available VR HMDs (like the the Oculus Rift S used in our experiments).
VibroWeight is effective in augmented haptic sensations of virtual objects, resulting in enhanced 
\textit{realism} and \textit{comfort}.

The contribution of this research is twofold. (1) The implementation and evaluation of a prototype, VibroWeight, that is based on low-cost materials and can be used to provide both haptic and vibration feedback during user-VR interaction. (2) A first exploration of liquid metal as an alternative material (e.g., against water) to provide fluid-based passive haptic feedback dynamically in VR environments. Our two user studies show that by leveraging both vibration feedback and fluid-based passive haptic feedback, VibroWeight can emulate changes in weights and center of gravity of virtual entities with detailed granularity
in VR environments.

\section{Related Work}

\subsection{Vibration Force Feedback}

Kim et al. \cite{kim2017study} proposed a portable haptic device that sends vibration and heat to the fingertips. The user feels the size of a virtual object through the intensity of the device's vibrations. Wu et al. \cite{wu2017virtual} used microspeakers to create simulated vibrations to emulate key touch feedback in a virtual environment. Choi et al. \cite{choi2017grabity} proposed Grabity, a device that allows users to feel different magnitudes of weight and force by exerting various amplitudes of asymmetric vibrations. Miyakami et al. developed Hapballoon \cite{miyakami2019hapballoon}, by which users can achieve improved touch and grasp of virtual objects with the bi-modal feedback of vibrations and temperature.
PseudoBend \cite{heo2019pseudobend} leverages a single 6-DOF force sensor and a vibrotactile actuator to render particle vibrations during object deformation, through changing force levels and torques, specifically for object deformation.
Pezent et al. \cite{pezent2019tasbi} proposed Tasbi, a multisensory haptic wristband that combines squeeze vibration and pseudo-tactile techniques. White et al. \cite{white2019case} combine vibration feedback with a weighted baseball prop to enhance the realism of virtual 
games. 

The above prior studies have some limitations, such as (1) very narrow application scenarios \cite{kim2017study,white2019case}, (2) noisy device 
\cite{choi2017grabity}, and (3) the simulated vibration feedback does not 
correspond well with virtual objects \cite{choi2020augmenting}. 
Inspired by~\cite{wu2017virtual} and~\cite{heo2019pseudobend}, we have been exploring alternative mechanisms of haptic feedback, which can work together with the most typical feedback driven by vibration signals~\cite{kim2017study,choi2020augmenting,miyakami2019hapballoon}. Our approach is based on fluid-driven haptic and, to our knowledge, is the first to use liquid metal to provide dynamic changes in the weight and center of gravity of virtual objects.

\subsection{Force Feedback for Weight Simulation}

Force being exerted on the human skin is an alternative solution to offer an illusion of weight to users. As shown in prior work~\cite{minamizawa2007gravity, stellmacher2021haptic}, external force from such devices on the user's fingers can stretch the skin surface, and the sight deformation of the skin produces an illusion of the object's weighting. However, such haptic simulators may not be practical and could be expensive to produce a wider range of user interactions in VR. For instance, in~\cite{minamizawa2007gravity}, the emulation of full-hand grasping of 3D shapes would need a full array of sensors on the user's palm. Another device, PuPoP~\cite{teng2018pupop}, utilizes an air pump to inflate an airbag. As such, users with PuPoP can sense the shape of virtual entities through the expanded airbag. However, PuPoP does not provide gravity as user feedback cues.

On the other hand, the changes in users' muscles when interacting with virtual objects make them more realistic, and design strategies that elicit muscle changes include on-device fixed weights matching those of tangible objects \cite{white2019case}, and user-driven stretched-and-flexed muscles detected by electromyography~\cite{muscle-haptic}. Additionally, our exhaustive search indicates that other recent solutions, such as Shifty~\cite{zenner2017shifty}, 
Aero-plane~\cite{je2019aero},
Drag:on~\cite{zenner2019drag},
SWISH~\cite{sagheb2019swish} and 
Transcalibur~\cite{shigeyama2018transcalibur}, leverage weight, as well as the center of gravity, to emulate weighting sensations of virtual entities. However, these solutions cannot adapt their absolute mass to various scenarios in VR, i.e., they only serve one particular purpose or a specific scenario. 

Instead of employing fixed absolute mass, liquid can serve as a feasible alternative to change the absolute weights of the simulating devices. Liquid is a flexible material and by controlling its flow in and out of a receptacle, it is possible to control its volume. One example of this type of device is, for instance, GravityCup~\cite{cheng2018gravitycup}. However, fluid-driven solutions in current devices have three key drawbacks: (1) they cannot simulate changes in the center of gravity of objects~\cite{cheng2018gravitycup, niiyama2014weight}, (2) the noise generated due to fluid movements~\cite{heo2018thor,je2019aero,zenner2019drag}, and (3) a delayed response to user actions leading to deteriorated or inferior user experience and immersion levels~\cite{je2019aero,ban2019directional}. As we show in our device and user experiments, the above drawbacks can be improved by employing liquid-state metal materials instead of water~\cite{cheng2018gravitycup}. 

Of the two common types of haptic devices, vibratory and fixed-weight, the former can be used to provide a sense of gravity shifts of objects but it is not useful for providing realistic feelings of the objects' weights~\cite{zenner2017shifty}. On the other hand, fixed-weight devices~\cite{white2019case} cannot adapt to a wider range of VR scenarios. Similarly, it is important to note that fluid-driven haptic devices can handle weight adjustments but not gravity shift. When designing VibroWeight we looked into the possibility of laying the groundwork for both \textit{adaptive weight changes and gravity shifts through bimodal cues driven by both vibration and force feedback}. Remarkably, VibroWeight is \textit{a first case that explores the use of liquid-state metal, which leads to a better performance (than water) and help reinforce our understanding of such bimodal feedback cues in gaming and non-gaming scenarios in VR}.

\section{Methodology}

This section explains our testbed virtual environment, hardware prototype, vibration and weight feedback, as well as system control. 

\subsection{Virtual Environment (VE)} 

\begin{figure}[h]
  \centering
  \includegraphics[width=.9\linewidth]{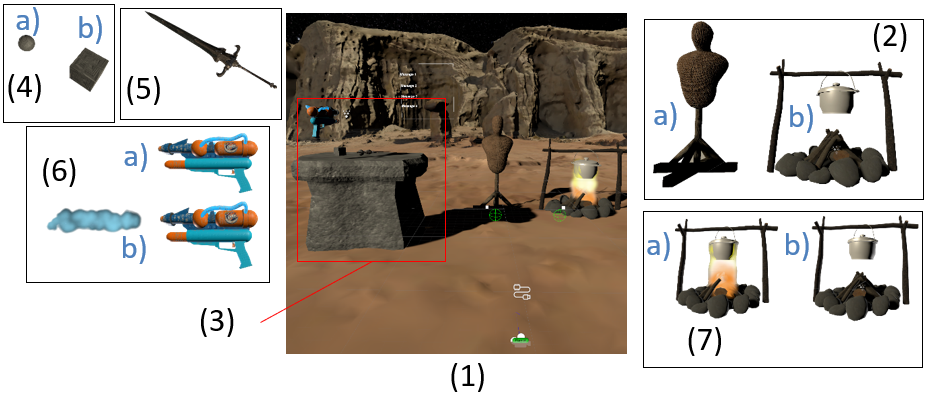}
  \caption{(1) The overview of the VE; (2) Game area, from left to right: a) dummy, b) fireplace; (3) Stone table with 4 items; (4) Stones, from left to right: (a) ball shaped stone, (b) square shaped stone; (5) Sword; (6) Water gun, from top to bottom: (a) not spraying water, (b) spraying water; (7) Fire place, top to bottom: (a) with fire, (b) without fire.}
  \label{fig:VEs}
\end{figure}

We built our VE with Unity3D. As shown in Fig.~\ref{fig:VEs}, the VE consisted of a stone table and four virtual objects (VOs) representing different weights. The objects' weight was proportional to their size and composition materials, and were meant to resemble real-life objects. Accordingly, users with VibroWeight could interact with these VOs and engage with other objects like the dummy (Fig.~\ref{fig:VEs}.(2)) and fireplace (Fig.~\ref{fig:VEs}.(7)).

\subsection{Simulating Weight with VibroWeight}

\begin{figure}[h]
  \centering
  \includegraphics[width=0.8\linewidth]{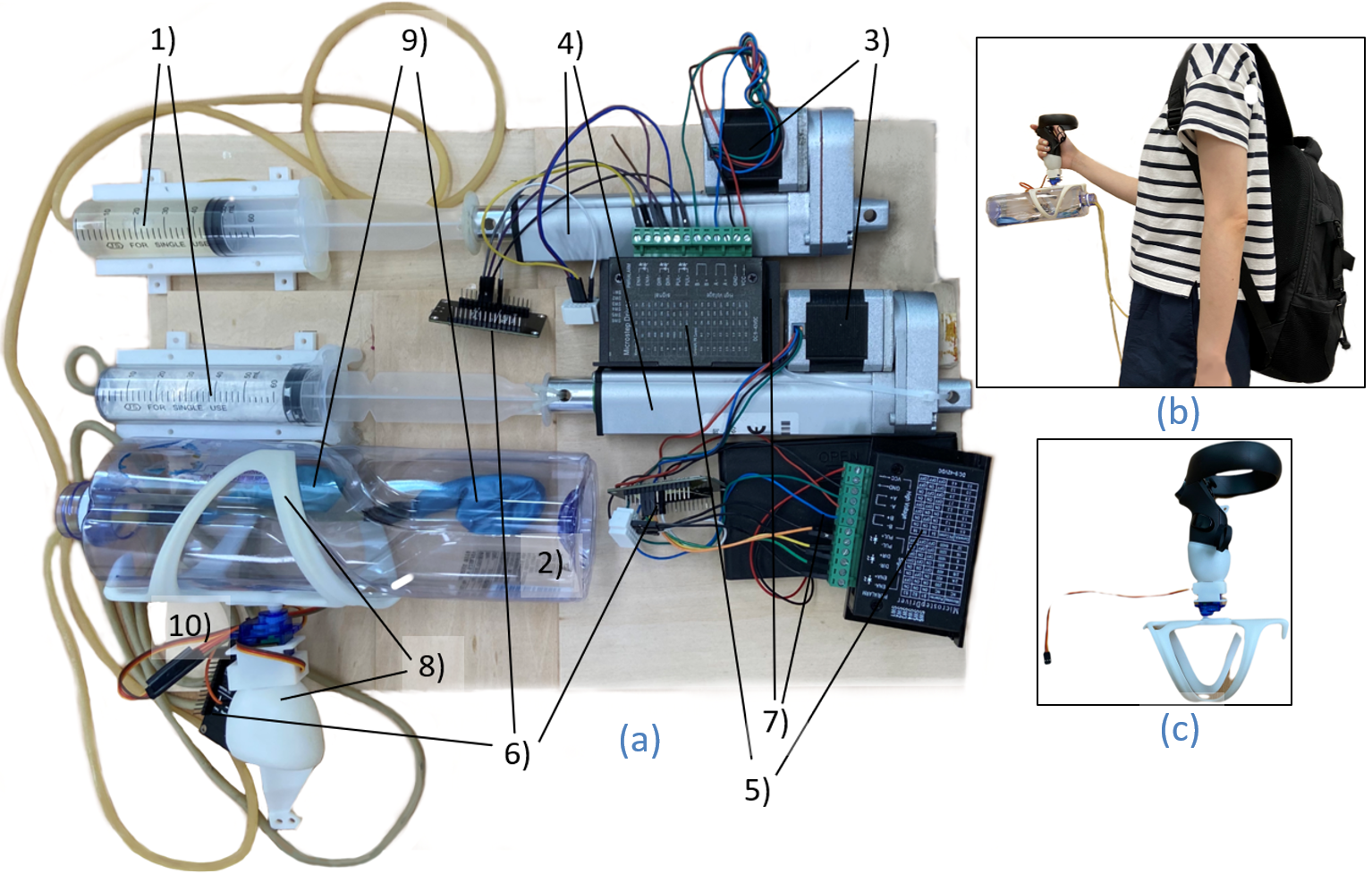}
  \caption{
  \textcolor{black}{(a)} Components of the VibroWeight prototype: (1) syringes, (2) bottle, (3) 42-step motors, (4) pushrod, (5) 9-42VDC encoder, (6) ESP8266s, (7) battery, (8) 3D-printed connector, (9) balloons as liquid receptacles, and (10) servo motor\textcolor{black}{; (b) A user with VibroWeight, components other than (2)\&(8-10) can be loaded into the backpack}\textcolor{black}{; (c) Detailed view of component \#8, the piece that connects the haptic device to the VR handheld controller.}}
  \label{fig:comp}
\end{figure}

\subsubsection{Hardware Prototype}
Fig.~\ref{fig:comp} depicts the parts and components of VibroWeight and the connections among the components. It also shows a user holding a VR controller with the haptic device attached to it. Component \#8 was modeled using 3DMAX and printed with a SLA 3D printing material (i.e., photosensitive resin) and can be adpated to other types of VR controllers. All parts and components feature good heat-resistance, have stable structure, and are light-weight (130.0 grams (g) before injecting the fluids). The structure can afford up to 902.8 g.

\subsubsection{User Interaction and Fluid Control}
\begin{figure}[h]
  \centering
  \includegraphics[width=\linewidth]{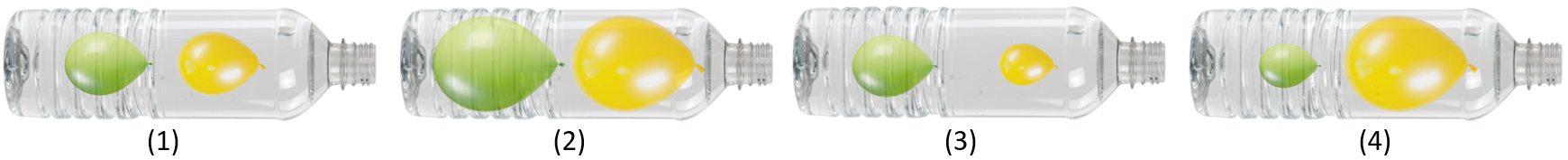}
  \caption{Schematic diagram of simulated weight and center of gravity changes: the balloons are filled with different volumes of fluid to allow sensing these changes. (1) Simulates light objects whereas (2) 
  heavy objects. (3) The object's center of mass is on the side away from the hand whereas (4) the object's center of mass is on the side close to the hand.}
  \label{fig:simulation}
\end{figure}

When the user picks up a VO in the VE, the fluid is then injected into the balloons by the pushrod at a speed of 60mm/s. The pushrod is controlled by a 42 stepper motor and a 9-42VDC encoder. 
Vibration feedback is realized by 
a small servo MG90 motor. 
The single-chip microcomputer ESP8266 controls the decoder and the servo to communicate with the computer that drives the VR HMD through Wi-Fi. Considering that user interaction with varied VOs can trigger different weight simulations, the fluid is injected into the separate balloons (Fig.~\ref{fig:simulation}) to emulate weight distributions of VOs and changes in their center of gravity. As our results show, this design works well to give users realistic feelings.

\begin{figure}[h]
  \centering
  \includegraphics[width=.75\linewidth]{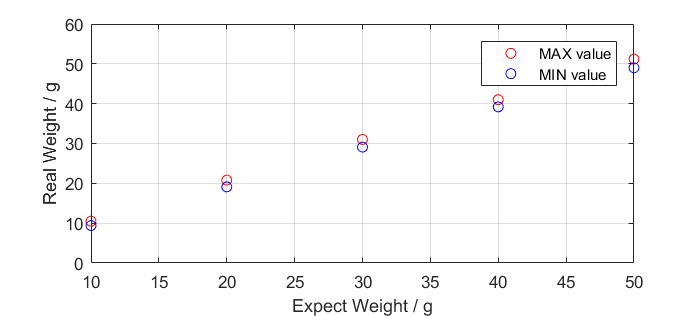}
  \caption{Measurements of VibroWeight's stability. Each specific weight was repeated 60 times.}
  \label{fig:robust-result}
\end{figure}

Since VibroWeight uses open-loop fluid control, we evaluated its robustness as follows. We defined five targeted weights (10g, 20g, 30g, 40g, and 50g) and measured the deviation from the target with 60 repetitions (i.e., 5 targets $\times$60 repetitions). Measurements were made using an electronic scale with an accuracy of 0.1g, and the weight measured was the weight of the equipment held in the user's hand minus the weight of the empty device. The weight of the empty device was measured separately.
Fig.~\ref{fig:robust-result} shows that VibroWeight was able to maintain a stable performance, with a deviated weight of just 4.4\%. Heavier weight of 50g resulted in higher erroneous fluid injection but no more than 2g (out of 50g).

\subsubsection{Liquid Materials}\label{sssec:metalliq}
VibroWeight's fluid materials were tested with both water and Galinstan, as part of a follow-up user experiment (Experiment B). 
The two fluid materials were chosen because water is easily accessible and commonly used in VR weight simulations \cite{cheng2018gravitycup}. Galinstan was employed to be compared with water because we wanted to find an alternative that could improve performance further. It is an eutectic alloy composed of 68.5\% Ga, 21.5\% In and 10\% Sn (by weight) \cite{liu2011characterization}. The melting point of this metal is -19$^\circ$ $C$ and it conserves its liquid state at room temperature~\cite{liu2011characterization}. The viscosity of Galinstan is $24 \times 10^{-4} \cdot P a \cdot s$, which is less than \textit{milk}, while its density is 6.44 (5.44 times greater than water). 
These two properties are important and indicate that Galinstan can maintain extremely high fluidity even under pressure. Remarkably, Galinstan is non-toxic~\cite{liu2011characterization} and hence guarantees user safety in case of fluid leakage.

\subsection{Vibration Feedback}\label{sssec:vibrate}
\begin{figure}[h]
  \centering
  \includegraphics[width=0.8\linewidth]{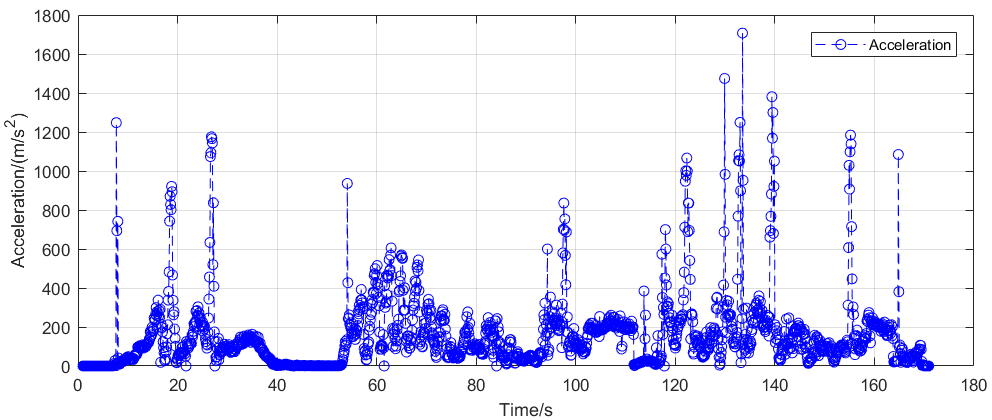}
  \caption{The acceleration of one of the participants while swinging a sword left-right in the VE. When the acceleration reaches 800, the oscillatory feedback is triggered.}
  \label{fig:accelerate}
\end{figure}

The weight feedback works in complement with vibration feedback, for instance, simulating a cylinder-shaped object in the real world which has both weight and center of gravity changes. Eq.~\ref{maths:eq1} depicts the law of motion that describes the device vibrations according to  the relationship between mass and moving acceleration. Eq.~\ref{maths:eq2} describes \textit{Damping Sine Wave} to deliver the sense of gravity shift of objects in the VE, e.g., a rod that hits a virtual object. VibroWeight triggers the vibration as oscillatory feedback, when the acceleration value of VOs exceeds the threshold of 800 \textcolor{black}{$\mathrm{m} / \mathrm{s}^{2}$}. Fig.~\ref{fig:accelerate} illustrates the acceleration and feedback cues when a sword in the VE is swung from one side to another. 

\begin{table}[h]
\tiny
  \caption{Symbols used in the formulas}
  \label{tab:symb}
 \resizebox{\columnwidth}{!}{ 
  \begin{tabular}{cl|cl}
    \toprule
    Symbol & Meaning & Symbol & Meaning \\
    \hline
    $F$ & Volume & $A$ & Plunger radius\\
    $m$ & Mass & $\lambda$ & Decay constant\\
    $a$ & Density & $\phi$ & Phase angle at some arbitrary point\\
    $y(t)$ & Height or plunger distance & $\omega$ & Angular Frequency\\
    \bottomrule
\end{tabular}}
\end{table}

\begin{equation}\label{maths:eq1}
\begin{array}{l}
F=m a
\end{array}
\end{equation}
\begin{equation}\label{maths:eq2}
\begin{array}{l}
y(t)=A \cdot e^{-\lambda t} \cdot(\cos (\omega t+\phi)+\sin (\omega t+\phi))
\end{array}\end{equation}

\subsection{System Control}
Figure 6 shows the control logic of the bimodal feedback system. ESP8266 receives user interaction via a Wi-Fi receiver and controls the hardware components such as the servo and stepper motors that control the fluid volume in and out of the device. Our application run in \textit{Unity} retrieves the data of user interaction with VOs in the VE, and makes real-time changes in weight and vibration accordingly. For example, when a user picks up a sword in the VE, \textit{Unity} can detect this information and send the message "SwordPickup" to ESP8266, which then with its \textit{Arduino} board processes this information and controls the pushrod to inject a specific amount of liquid into the balloons to adjust the device's weight.

\begin{figure}[h]
  \centering
  \includegraphics[width=.75\linewidth]{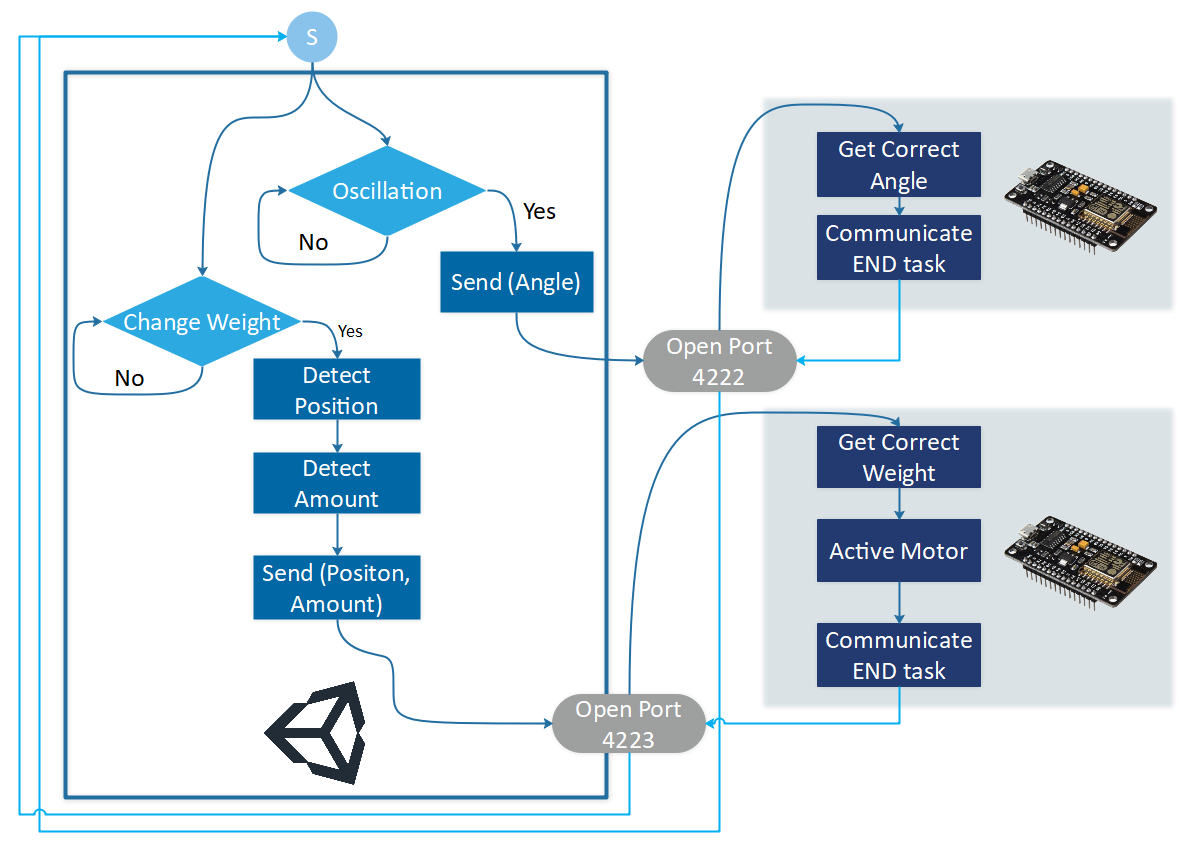}
  \caption{An overview of the system control for bimodal feedback. The microcontroller ESP8266 detects 
  the current state and activates the transfer mechanism. The Unity application informs the controller regarding the anticipated changes of weight and vibration information.}
  \label{fig:SystemOverview}
\end{figure}

\section{User Evaluation}
To explore whether vibroWeight is effective and if Galinstan is a viable alternative liquid-state material, we conducted two experiments to understand users' perception of VibroWeight with two liquid settings: (1) Water and (2) Galinstan (Sec.~\ref{sssec:metalliq}).
In both experiments, we evaluated three types of controllers: (1) a typical commercial VR controller (which in our case is the one for the Oculus Rift S and represents the baseline condition); (2) VibroWeight without oscillating feedback; and (3) VibroWeight with oscillating feedback (i.e., vibration, see Section~\ref{sssec:vibrate}). The participants had to complete two tasks in two scenarios, namely \textit{No Game} and \textit{Game}. The two tasks serve to examine whether VibroWeight could improve the sense of presence in both gaming or non-gaming scenarios in VR. 
In \textit{No Game}, participants only interacted with the four VOs on the table (Fig.~\ref{fig:VEs}). In \textit{Game}, participants threw \textit{stones} or swung a \textit{sword} to attack the dummy and used a \textit{water gun} to extinguish a fire in the fireplace. In other words, six conditions (2 \textit{Liquid} Types $\times$ 3 \textit{Controller} Types) were assigned to each participant.
We applied Latin Square design to counterbalance the conditions and mitigate any carry-over effects. 
Participants had to fill in a questionnaire after each condition (6 questionnaires in total). As shown in~Table \ref{tab:questionnatire}, three qualitative metrics were used to measure the sense of presence, namely \textit{Realism}, \textit{Comfort} and \textit{Enjoyment}. An identical user questionnaire was employed in the two experiments to ensure consistency in the data collection.

\begin{table}[h]
\tiny
  \caption{Questionnaire used in the experiment (5-point Likert scale)}
  \label{tab:questionnatire}
 \resizebox{\columnwidth}{!}{ 
  \begin{tabular}{cc}
    \toprule
     Metrics & Questions\\
    \hline
    \multirow{6}*{Realism}
    & I felt like I was holding the same object I was seeing\\
    & I forgot I was holding a controller\\
    & Even if I had my eyes closed, I could imagine the virtual object I was holding\\
    & I was concerned the virtual object could affect my body\\ 
    & I was concerned I could break the virtual object \\
    & I was concerned the virtual object could affect others around me\\
    \cline{2-0}
    \multirow{2}*{Comfort}
    & My arms were tired; I was tired in general\\
    & The device was heavy/loud; I was bothered by the device\\
    \cline{2-0}
    \multirow{3}*{Enjoyment}
    & Using the device was really enjoyable; The VE was enjoyable\\
    & The controller and device really added to the experience\\
    & I wish I could keep using the device in other VE; I could feel changes in the VOs\\
    \bottomrule
\end{tabular}}
\end{table}

The study aimed to examine the following:
(1) whether VibroWeight could simulate weight and center of gravity changes;
(2) whether VibroWeight with oscillating feedback could enhance user experience;
(3) whether Galinstan as the fluid could provide better simulation and performance than water;
(4) the effectiveness and feasibility of integrating VibroWeight into VR gaming and non-gaming scenarios.
To this end, we recruited two groups of participants from a local university for the experiment (A: with Water) [participant number = 16 (10M /6F),
\textcolor{black}{$\overline{age}$ = 22.06,} 
age-range = 19 -- 29] and (B: with Galinstan) [participant number = 13 (6M /7F), 
\textcolor{black}{$\overline{age}$ = 21.69,} 
age-range = 20 -- 28]. None of them reported a history of physical or muscular discomfort. They had normal or corrected-to-normal vision. 82.76\% of them had some VR experiences.
Their participation was entirely voluntary and based on informed consent. The study falls under the low risk research category. It was reviewed and approved by the University Ethics Committee at Xi'an Jiaotong-Liverpool University. 

\subsection{Experiment A: Water as the Liquid Material}
\subsubsection{Comparison of the three controllers}

\begin{figure}[h]
  \centering
  \includegraphics[width=.8\linewidth]{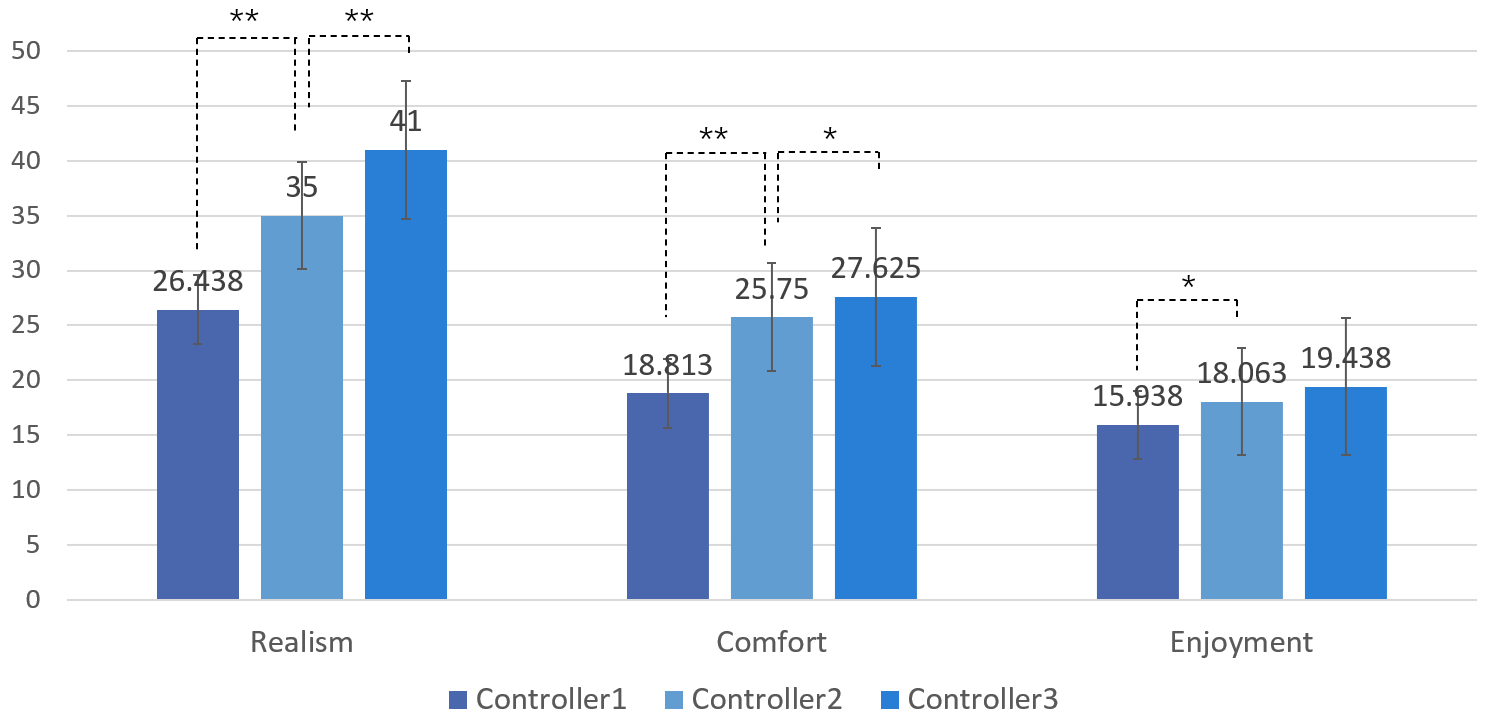}
  \caption{\textcolor{black}{The average scores of the three controllers with water setting in the three aspects of Realism, Comfort and Enjoyment; (Controller 1: commercial VR controller; Controller 2: VibroWeight without oscillating feedback; Controller 3: VibroWeight with oscillating feedback); asterisk indicate statistically significant differences (p $<$ .05 ($\ast$)\textcolor{black}{, p $<$ .01 ($\ast\ast$)}).}}
  \label{fig:ex1-controllers}
\end{figure}

The data were analyzed using both statistical inference methods and data visualizations. First, we conducted a Shapiro-Wilk test to check the normality of the data. As all data were classified as normally distributed, they received a parametric analysis. Also, we conducted Mauchly's Sphericity Test to verify if the assumption of sphericity had been violated. Finally, we ran Repeated Measures ANOVA (RM-ANOVA) with Bonferroni correction to detect significant main effects. When the assumption of sphericity was violated, we employed the Greenhouse-Geisser correction to adjust the Degrees of Freedom. 

Figure~\ref{fig:ex1-controllers} summarizes the main results of participants' perceptions of VibroWeight with water according to the three controllers. RM-ANOVA indicated statistical significance in \textit{Realism} among the three controllers (F$_{1.779, 26.682}$ = 28.213, p $<$ .0005), and post-hoc analysis using Bonferroni correction showed that VibroWeight without oscillatory feedback, i.e., Controller 2, significantly outperformed the baseline controller, Controller 1 (p = .003). It also showed that VibroWeight with oscillatory feedback, i.e., Controller 3, significantly outperformed its counterpart without oscillatory feedback, i.e., Controller 2 (p = .005).

Regarding \textit{Comfort}, statistical significance was found among the three controllers
(F$_{1.387, 20.805}$ = 30.875, p $<$ .0005). Participants indicated that (1) Controller 2 led to a higher level of comfort than Controller 1 (p $<$ .0005) and (2) Controller 3 outweighed Controller 2 (p = .049). 
As for \textit{Enjoyment}, we found a main effect (F$_{1.640, 24.602}$ = 10.573, p = .001) among the three controllers (Controller 3 $>$ 2 (p = .098, but not significant) and Controllers 2 $>$ 1 (p = .047)). 
The above findings show that, in general, VibroWeight (using water) offers a better sense of presence to users, in terms of all the three metrics. The additional vibration feedback further enhanced the user experience of VibroWeight (than without it, especially for Realism and Comfort).

\begin{figure}[h]
  \centering
  \includegraphics[width=\linewidth]{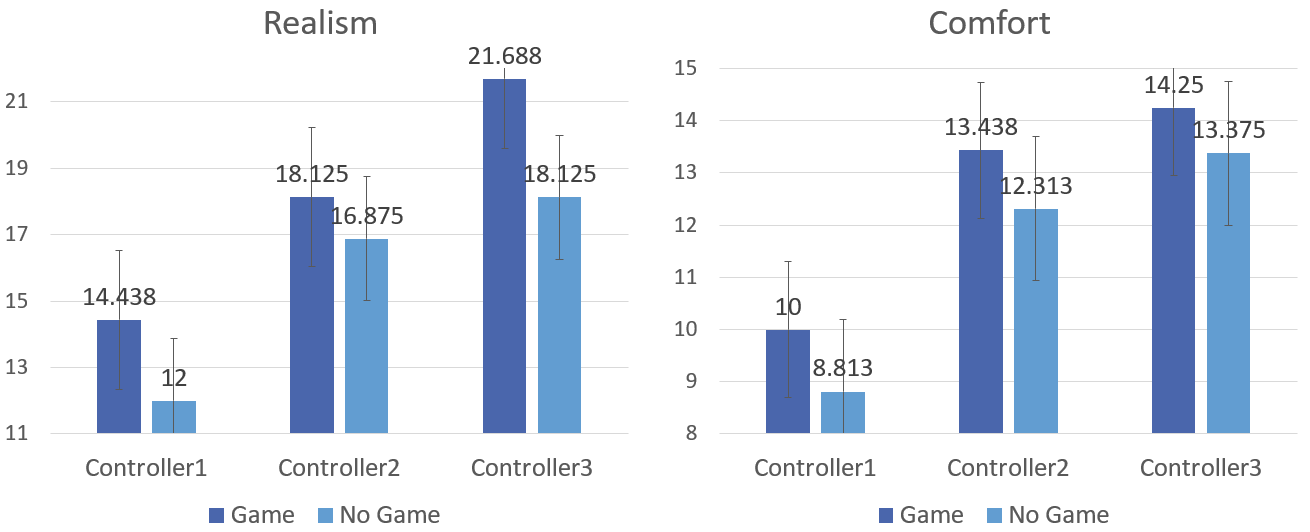}
    \caption{The average scores of the three controllers (using water) compared in \textit{NO Game} (only interacting with the VOs but not within a game setting) and \textit{Game} (playing game-like activities with the VOs, e.g., using the sword, stones to interact with dummy, and water gun to extinguish fire) conditions in the 
    \textcolor{black}{two aspects of Realism and Comfort.}}
  \label{fig:ex1-separate}
\end{figure}

\subsubsection{Impact of game interaction}
As illustrated in Fig.~\ref{fig:ex1-separate}, we ran RM-ANOVA to analyze two scenarios: \textit{No Game} and \textit{Game}, and found significant main differences in Realism 
(F$_{1, 15}$ = 7.559, p = .015) but not in Comfort 
(F$_{1, 15}$ = 1.947, p = .183) between the two (game/no game) conditions. It is important to note that the average scores of \textit{Game} task(s) are slightly higher than their counterpart of \textit{NO Game}. In line with previous work~\cite{krcmar2011effects}, participants with VibroWeight in gaming scenarios feel a strong sense of \textit{realism} than in the non-gaming situations.

\subsection{Experiment B: Metal (Galinstan) as the Liquid Material}
\subsubsection{Comparison of the three controllers}

\begin{figure}[h]
  \centering
  \includegraphics[width=.8\linewidth]{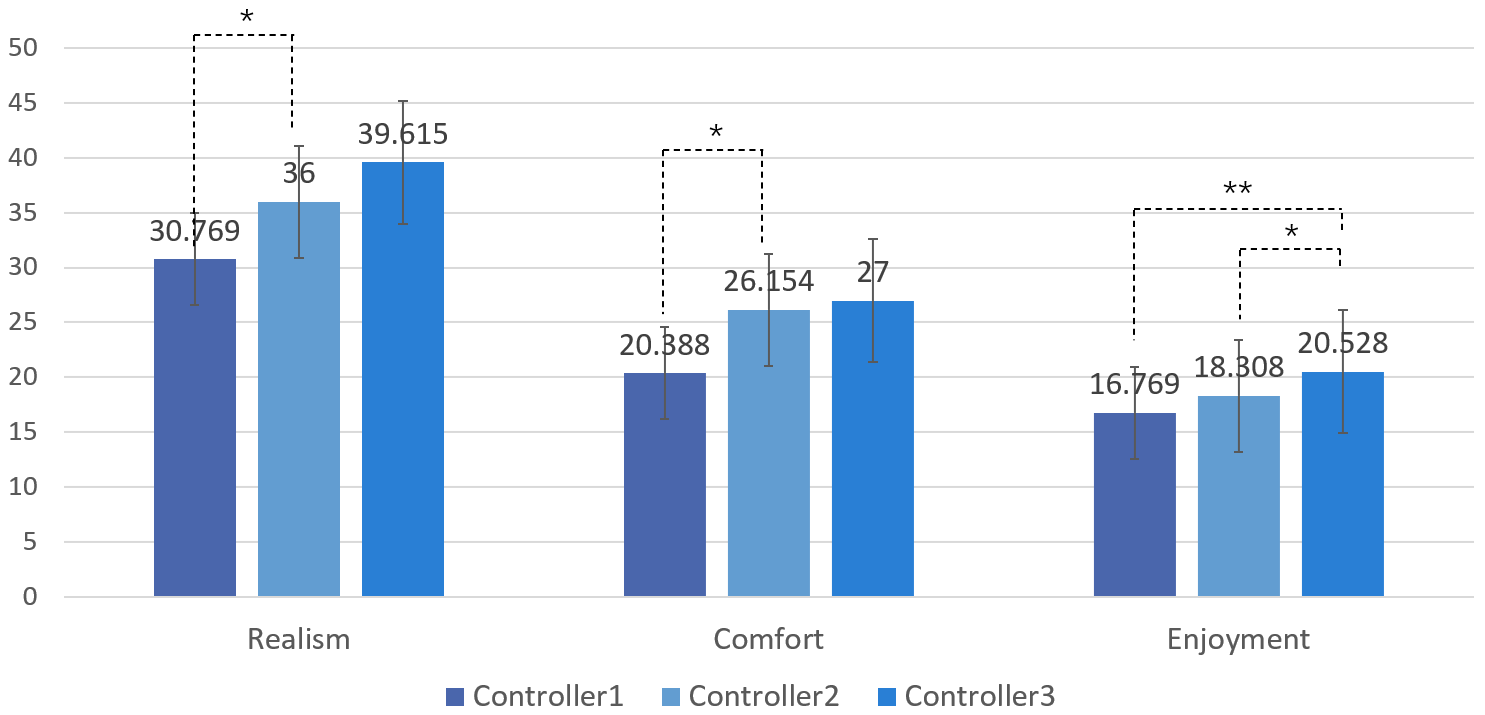}
    \caption{\textcolor{black}{The average scores of the three controllers in the Galinstan setting with the same three aspects as in Experiment A; asterisk indicate statistically significant differences (p $<$ .05 ($\ast$)\textcolor{black}{, p $<$ .01 ($\ast\ast$)}).}}
  \label{fig:exp2-controllers}
\end{figure}

We ran RM-ANOVA with a Greenhouse-Geisser correction and found statistically significant differences in all three metrics: \textit{Realism} 
\textcolor{black}{(F$_{1.212, 14.542}$ = 8.940, p = .007)}, \textit{Comfort} 
\textcolor{black}{(F$_{1.104, 13.252}$ = 11.201, p = .004)}, and \textit{Enjoyment} \textcolor{black}{(F$_{1.641, 19.696}$ = 7.248, p = .006).}
As Figure~\ref{fig:exp2-controllers} shows, VibroWeight with Galinstan led participants to enhanced user experiences in all metrics.
Similar to Experiment A, VibroWeight now with Galinstan achieved the same results for all three metrics: Controller 3 $>$ Controller 2 $>$ Controller 1. In other words, Controller 2 brought significantly higher \textit{Realism} and \textit{Comfort} than Controller 1 (both p = .029), while Controller 3 (with vibration) led to significant enhancement in \textit{Enjoyment} over the baseline controller (Controller 1, p = .010) and Controller 2 (VibroWeight without vibration, p = .033). 

\begin{figure}[h]
  \centering
  \includegraphics[width=\linewidth]{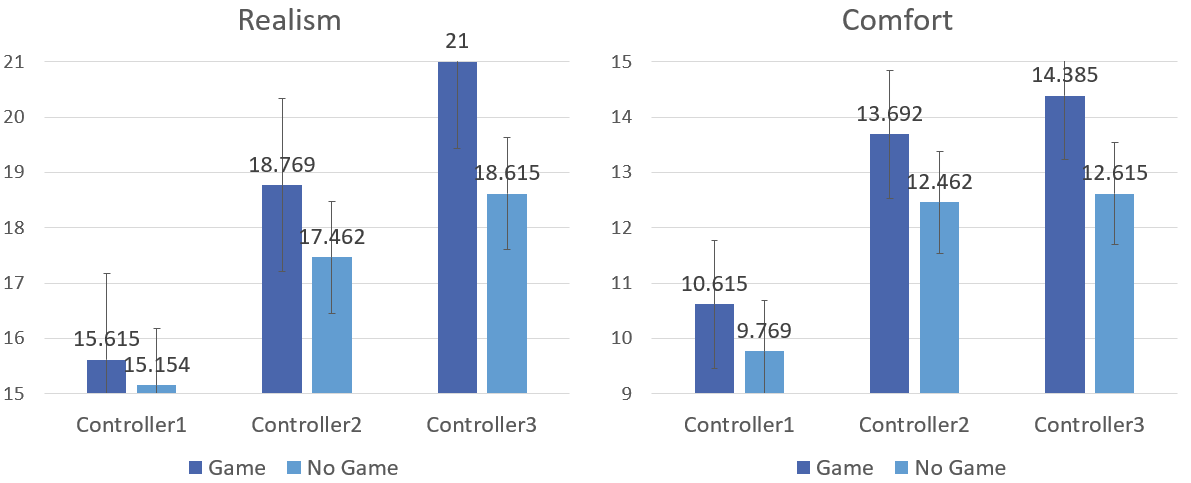}
    \caption{\textcolor{black}{Comparison of the average scores of the three controllers in the Galinstan experiment for the same 
    \textcolor{black}{two}
    aspects as in Experiment 1 in the \textit{NO game} and \textit{Game} conditions.}}
  \label{fig:exp2-game}
\end{figure}

\subsubsection{Impact of game interaction}
Using the same approach as in Experiment 1, RM-ANOVA showed a significant effect on \textit{Comfort} 
F$_{1,12}$ = 9.709, p = .009) but not on \textit{Realism} 
(F$_{1,12}$ = 2.782, p = .121). Figure~\ref{fig:exp2-game} shows the results of participants' perceptions under both \textit{No Game} and \textit{Game} scenarios. 

\subsection{A Comparison: Water vs. Metal}

\begin{figure}[h]
  \centering
  \includegraphics[width=\linewidth]{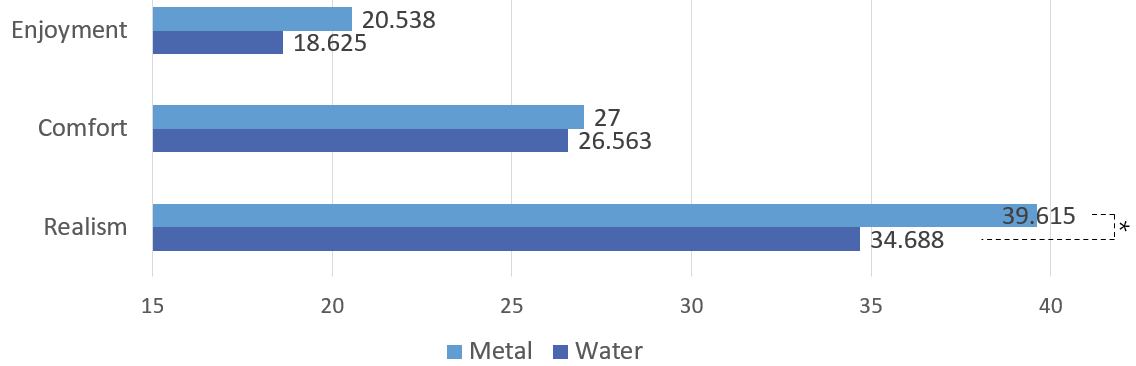}
  \caption{Estimated Marginal Means for water and liquid metal.}
  \label{fig:liqliq} 
\end{figure}

We further examined the metrics in Experiments 1 and 2 with the two types of liquids. We ran a Mann-Whitney test to check the effect of fluid type on participants' experience. As shown in Fig.~\ref{fig:liqliq}, the three metrics under \textit{Metal} generally achieved higher mean values than \textit{Water}. We also found a significant difference in \textit{Realism} in particular (U = 61.5, p = .031 (1-tailed)) between the \textit{Metal} and \textit{Water}. 

\subsection{Key Implications and Limitations}
Our research has presented a viable hardware prototype which uses a fluid-driven system and vibration haptic feedback to enable users to sense weight and center of gravity changes of virtual objects and thus a higher sense of presence. Our prototype has led to comfort metrics that are higher than a typical commercial VR controller. In general, VibroWeight outperforms a typical controller and is helpful for both gaming and non-gaming scenarios. Including oscillatory feedback further enhances its usefulness while using liquid metal seems to have provided a greater sense of realism than with water. Our tailor-made metal fluid, Galinstan composed of monomers of gallium, indium and tin, can be easily purchased online at a low price of \$0.3 USD/g, which is cheaper than similar compounds used by other researchers (e.g., \$0.6 USD/g in ~\cite{niiyama2014weight}). As such, metal-state liquids represent viable alternatives for fluid-driven haptic devices. 

Regarding the limitations of this work, our focus is on low-cost materials and, if cost is not an issue, other smaller components can be explored to make the overall device smaller and lighter so that it can be more mobile. In addition, we have not explored the use of VibroWeight in other types of VR games and applications where objects' weight and oscillatory feedback is useful, for example when lifting objects in an authentication system~\cite{olade2021BioMove}, holding different types of small excavation tools in an exploratory game~\cite{Liu2021.VRRelic}, or manipulating 3D objects in a modelling environment~\cite{Yu2021Gaze3D}. While it is possible to simulate the weight and center of gravity changes of such virtual objects, further research is needed to assess the performance, stability, and suitability of haptic devices like VibroWeight in these VR scenarios. 

\section{Conclusion}
This paper presented our explorations of a new hardware prototype, VibroWeight, that leverages both oscillatory vibration and liquid based force haptic to provide bimodal feedback cues for weight and center of gravity changes of objects in virtual reality (VR). Our experiments show that users with VibroWeight can achieve an enhanced sense of presence in virtual environments, especially for realism and comfort. Our device can be developed using low-cost materials and can be integrated in a typical VR controller, like the Oculus. Our design is the first to show the practicality of using liquid metal in a haptic device for VR with positive results.

\section*{Acknowledgment}
The authors want to thank the participants for their time and the reviewers for their insightful comments. This work was supported in part by Xi'an Jiaotong-Liverpool University Key Special Fund (\#KSF-A-03).

\bibliographystyle{IEEEtran}
\bibliography{new}
\end{document}